\documentclass[journal]{IEEEtran}
\ifCLASSINFOpdf
  \usepackage[pdftex]{graphicx}
\else
  \usepackage[dvips]{graphicx}
\fi
\usepackage{amsmath}
\usepackage{algorithm}
\usepackage{algorithmic}

\usepackage{cleveref}

\usepackage{pgfplots}
\usepackage{siunitx}
\usepackage{tikz}

\pgfplotsset{compat=1.18}

\begin{document}
%
\title{Hourglass Sorting: A novel parallel sorting algorithm and its implementation}
%
%
%

\author{Daniel Báscones,
        Borja Morcillo
\thanks{Authors are with the Department
of Computer Architecture and Automatics, Universidad Complutense de Madrid, Madrid, Spain. e-mail: (danibasc@ucm.es).}}

\maketitle

\begin{abstract}
Sorting is one of the fundamental problems in computer science. Playing a role in many processes, it has a lower complexity bound imposed by $\mathcal{O}(n\log{n})$ when executing on a sequential machine. This limit can be brought down to sub-linear times thanks to parallelization techniques that increase the number of comparisons done in parallel. This, however, increases the cost of implementation, which limits the application of such techniques. Moreover, as the size of the arrays increases, a bottleneck arises in moving the vast quantities of data required at the input, and generated at the output of such sorter. This might impose time requirements much stricter than those of the sorting itself. In this paper, a novel parallel sorter is proposed for the specific case where the input is parallel, but the output is serial. The design is then implemented and verified on an FPGA within the context of a quantum LDPC decoder. A latency of $\log{n}$ is achieved for the output of the first element, after which the rest stream out for a total sorting time of $n+\log{n}$. Contrary to other parallel sorting methods, clock speed does not degrade with $n$, and resources scale linearly with input size.
\end{abstract}


%
\IEEEpeerreviewmaketitle

\section{Introduction}

\IEEEPARstart{S}{orting} algorithms are at the core of many of the processes that take part in computing. Appearing in databases, scheduling, data analytics, networking, recommendation systems, computational biology and, more recently, computer graphics or AI. Consequently, they are one of the most researched families of algorithms, and often the first to be studied in any basic algorithms course.

The formal proof that outlines the minimum $\mathcal{O}(n\cdot\operatorname{log}(n))$ complexity of sorting was published by D. Knuth \cite{knuth1998art}, along with a collection of algorithms that reach that bound. Merge sort (attributed to John von Neumann in 1945; however lacking formal publication), heap sort \cite{williams1964algorithm}, and radix sort \cite{hollerith1889art} (which appeared prior to modern computers being available) are some examples. Quicksort \cite{hoare1962quicksort}, which only ensures an \emph{average} complexity of $\mathcal{O}(n\cdot\operatorname{log}(n))$ is -perhaps surprisingly- the most used, due to its simplicity and in-place sorting capabilities.

As time progressed, and data sets enlarged, the need for faster sorting was fulfilled by going parallel. The first sorting networks \cite{batcher1968sorting} were able to sort an array in $\mathcal{O}(\operatorname{log}^2(n))$ time using $\mathcal{O}(n\cdot\operatorname{log}^2(n))$ resources. These were eventually refined \cite{ajtai19830} to $\mathcal{O}(\operatorname{log}(n))$ complexity and $\mathcal{O}(n\cdot\operatorname{log}(n))$ resources at the cost of a higher constant, which, unfortunately, made them impractical for most cases. Finally, it was proven that, given a CREW PRAM (Concurrent Read Exclusive Write Parallel Random Access Machine) architecture, a parallel merge sort algorithm \cite{cole1988parallel} can sort an array in $\mathcal{O}(\operatorname{log}(n))$ time using just $\mathcal{O}(n)$ processors.

In practice, hardware-accelerated sorting networks \cite{mueller2012sorting} are used when the problem is small and speed is critical. Parallel sorting techniques \cite{akl2014parallel} for larger datasets usually work on general-purpose hardware: they partition, sort, and merge the arrays using sequential sorting and clever data interleaving. 

There is, however, a gap in the case that faster than $\mathcal{O}(n\cdot\log (n))$ sorting is needed for large datasets: Hardware accelerated sorting networks become unfeasible due to resource use, and parallel techniques on general purpose hardware might not be sufficiently fast. Efforts have been made in bridging this gap, and two techniques shine above others: 1) Iterate over a small sorting network \cite{sklyarov2014high}. This reduces hardware use by a factor of $\log(n)$, increasing latency to linear time. 2) Trade off time and space complexity by selecting the input/output width and streaming data in chunks \cite{ortiz2010configurable}. Further explored by \cite{zuluaga2016streaming}, both exploit that only a fixed amount of data $w$ is processed in parallel, reducing hardware complexity, and increasing processing time, by a factor of $n/w$.

All of these algorithms (either serial or parallel) share a common characteristic: a fixed size $w$ for \emph{both} the input and output. However, it is not always that the processing is symmetrical. For certain applications, we might be interested in asymmetrical SIPO (Serial-In Parallel-Out) or PISO (Parallel-In Serial-Out) schemes. A SIPO scheme with incremental sorting capabilities is presented in \cite{ortiz2010configurable} with an application in statistical signal processing, leveraging the fact that a single datum arrives per cycle to achieve $\mathcal{O}(1)$ sorting time. 
While PISO schemes have not been found directly in the literature, any PIPO (Parallel In Parallel Out) algorithm (such as sorting networks) can be adapted by placing a shift-register at its output. The cost (in hardware) would thus be slightly more than the original, at least $\mathcal{O}(n\cdot \operatorname{log}(n))$ considering the smallest sorting networks \cite{ajtai19830}.

In this work, we propose a novel sorting method for the PISO scheme, that has $\mathcal{O}(n)$ hardware complexity, and is able to output its first sorted datum in $\mathcal{O}(\operatorname{log}(n))$ time. Furthermore, its clock cycle is independent of the input array size (a property not all sorting accelerators exhibit, that hinders many when scaling up), making it highly scalable. This method has been implemented in an FPGA, and is currently used in a BP-OSD (Belief Propagation - Ordered Statistics Decoder) \cite{jiang2007reliability} implementation as the intermediate step for ordering the outputs of BP.

\textbf{NOTE:} Throughout this paper, it is assumed for simplicity that ascending sorting (lowest first) is performed. All of the designs, however, can be adapted to descending order trivially.

\section{Problem Context}

In Quantum Computers, one of the most critical problems to solve is that of data stability. Qbits (Quantum Bits) degenerate very rapidly \cite{terhal2015quantum} and with error rates many orders of magnitude above classical bits. In the classical world, these errors are often mitigated with some sort of ECC \cite{peterson1972error} (Error Correction Code). In essence, more bits than necessary are used to represent the information, so that if one flips, others can be used to recover it.

In the context of quantum computing, LDPC (Low Density Parity Check) quantum codes are used. Given the high error rates inherent to current technologies, these codes often use several physical qbits to just represent a single logical qbit. When checking for errors, the information from many different parity checks is gathered in order to understand what failure might have arrived at that state. How this is done is beyond the scope of this paper, but the general workings of the algorithm are outlined below:

When \emph{solving} a LDPC, BP is applied first. This algorithm outputs a probability for each qbit, indicating if it experienced (or not) an error. In most cases and under low error rates, this algorithm alone can be 100\% confident in which qbits failed and which did not. This information is fed back to the system to correct the errors. In some cases, BP might not be sure of what caused the failure, in which case OSD is applied after. OSD requires the output of BP to be sorted by probability, and mathematically solves the system of parity check equations that involve the most likely qbits to have failed, finding a solution that satisfies them. This information can in turn be use to recover any errors.

To understand why sorting fast is necessary, we now analyze algorithm complexity as a function of three variables involved in the calculations: $n$, which is the number of potential failure points, $m; \  m\ll n$, which is the number of checks and $k; \ k < m$ which is the number of iterations that BP performs. BP works in $\mathcal{O}(n\cdot k)$ time, but can be fully parallelized over $n$ to work in $\mathcal{O}(k) < \mathcal{O}(m)$. OSD works in $\mathcal{O}(m^2n)$, and can be parallelized over $\mathcal{O}(m^2)$ to work in $\mathcal{O}(n)$. This would at first suggest that parallelizing BP is unnecessary, but it has been found experimentally that it is probabilistically enough to run OSD in $\mathcal{O}(m)$ by using only partial information from BP, which makes the latter's acceleration necessary.

With this in mind, both BP and OSD are working at $\mathcal{O}(m)$, with an array sorting of $n \gg m$ elements in between. Our target is to lower this to $\mathcal{O}(m)$ or below, to match the complexity of the OSD decoder. More specifically, it is sufficient to obtain the $m$ lowest values in  $\mathcal{O}(m)$, matching the rest of the process's complexity. Classical serial algorithms that work in $\mathcal{O}(n\cdot \operatorname{log}(n))$ will not be fast enough, while parallel algorithms that work in $\mathcal{O}(\operatorname{log}(n))$ time take resources proportional to at least $\mathcal{O}(n\cdot \operatorname{log}(n))$, which experimentally exceeds resource availability. Furthermore, the input to OSD is serial, so having a fully parallel output is unnecessary.

\section{Design Idea}\label{sec:idea}

Ideally, then, we strive for an algorithm that finds the lower $m$ values in $\mathcal{O}(m)$, with a hardware complexity of at most $\mathcal{O}(n)$. Note that, despite $m\ll n$, we find that $\operatorname{log}(n) < m$.

\begin{figure}[h]
  \centering
  \includegraphics[width=0.7\linewidth]{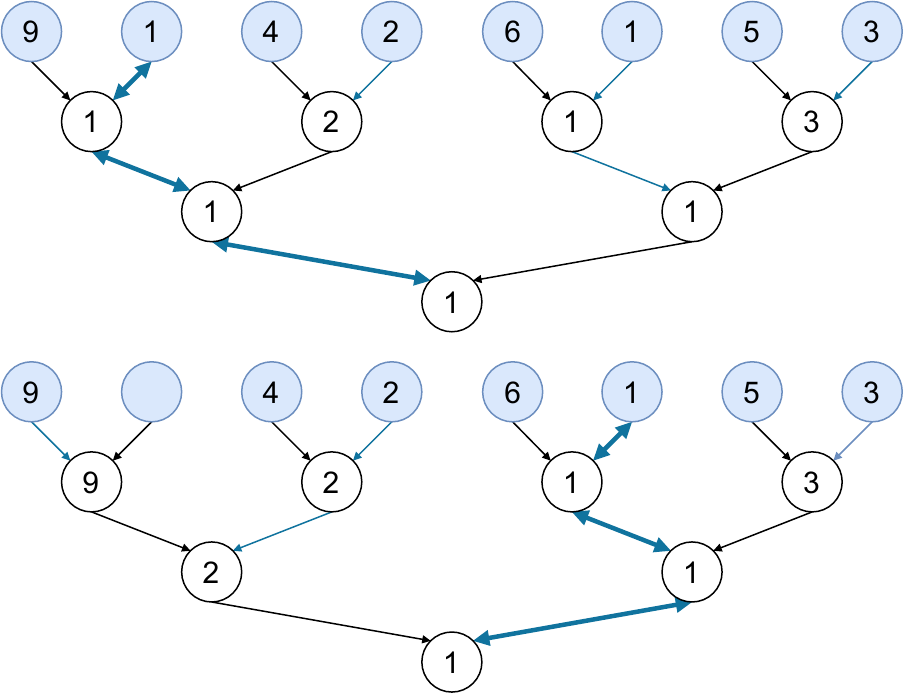}
  \caption{Two steps of a naive way of sorting which does not scale well with input size. Registers are shown as light blue, and only exist for the first layer.}
  \label{fig:naivetree}
\end{figure}

Given that our input (coming from BP) is parallel and our output (going to OSD) is serial, this maps very naturally to a tree-like structure (\Cref{fig:naivetree}) where the input gets progressively reduced to a single point. Indeed, if we just had a tree of comparators with no registers in between, we would find and delete the minimum value trivially as seen in \Cref{code:comptree}. 

\begin{algorithm}
\caption{Unregistered comparator tree algorithm}
\label{code:comptree}

\textbf{Input:} Values $D_1$, $D_2$ \\
\textbf{Input:} Valid flags $V_1$, $V_2$ \\
\textbf{Outputs:} $V_{out}$; $D_{out}$; \\
\textbf{Input:} Reset signal $R_{out}$ \\
\textbf{Outputs:} Reset signal $R_1, R_2$

\begin{algorithmic}[1]
\ENSURE Values output in ascending order; sorting is stable.
\FOR{Each triplet of nodes in the tree, in parallel}
    \STATE $V_{out} \leftarrow V_1$ \OR $V_2$
    \IF[both valid, select lowest] {$V_1$ \AND $V_2$}
        \STATE $D_{out} \leftarrow \min(D_1, D_2)$
        \STATE $R_{\operatorname{argmin}(R_1, R_2)} \leftarrow R_{out}$
    \ELSIF[only $V_1$ valid] {$V_1$}
        \STATE $D_{out},R_1 \leftarrow D_1,R_{out}$
    \ELSIF[only $V_2$ valid] {$V_2$}
        \STATE $D_{out},R_2 \leftarrow D_2,R_{out}$
    \ENDIF
    
\ENDFOR
\end{algorithmic}
\end{algorithm}

For this to work at a rate of a value per cycle, the lowest value has to travel through the tree of comparators to the bottom, going through $\log(n)$ of them. Furthermore, a reset signal $R$ has to then travel up the tree to delete/disable that value, preventing it from participating in further iterations. As such, even though at first it seems we achieve constant time, the critical path for this design would scale with $\mathcal{O}(\operatorname{log}(n))$, so extracting $m$ elements would result in a cost of $\mathcal{O}(\operatorname{log}(n))\cdot m$ time wise, an increase over our target of $\mathcal{O}(m)$.

A first approach to solving this stems from the fact that a signal has to travel back and forth between the root and leaf nodes, resulting in not only long critical paths, but also excessive fan-outs. A direct, yet naive solution appears to be just segmenting the levels in the tree by placing a register after each node. This creates a structure where isolated registered comparators compare the input from two registers, placing the lowest of both in an output register. Let's call this process an ``operation'' and consider that, for a registered comparator to perform an operation, the following conditions need to be satisfied:
\begin{itemize}
    \item Both input registers contain a valid value, or
    \item One input register is valid and the other empty (neither it nor any parent is valid)
    \item The output register is empty
\end{itemize}

This process is illustrated in \Cref{code:bubbly}. Each node waits for both inputs to be either valid or empty, and selects the minimum value to pass forward. This is conditioned to the output register being empty, in which case the value is moved down a level within the tree. When both inputs are empty (i.e: no more values will fall through them) and the output has been read, it is marked as empty, continuing the process below.

\begin{algorithm}
\caption{Registered comparator algorithm}
\label{code:bubbly}

\textbf{Input:} Data Registers $D_1$, $D_2$ \\
\textbf{Input:} Valid data flags $V_1$, $V_2$ \\
\textbf{Input:} Empty subtree flags $E_1$, $E_2$ \\
\textbf{Outputs:} $D_{out}$; $E_{out}$; $V_{out}$ 

\begin{algorithmic}[1]
\ENSURE $V_i \implies \neg E_i$
\ENSURE Values output in ascending order; sorting is stable.
\FOR{Each triplet of nodes in the tree, in parallel}
    \IF {Input registers belong to the first layer}
        \STATE $E_i \leftarrow \neg V_i$
    \ENDIF
    \IF {$V_1$ \AND $V_2$ \AND $\neg V_{out}$}
        \STATE $D_{out} \leftarrow \min(D_1, D_2)$
        \STATE $V_{\operatorname{argmin}(D_1, D_2)} \leftarrow \FALSE$
        \STATE $V_{out} \leftarrow \TRUE$
    \ELSIF {$V_1$ \AND $E_2$ \AND $\neg V_{out}$}
        \STATE $D_{out}, V_1, V_{out} \leftarrow D_1,\FALSE,\TRUE$
    \ELSIF {$V_2$ \AND $E_1$ \AND $\neg V_{out}$}
        \STATE $D_{out},V_2,V_{out} \leftarrow D_2,\FALSE,\TRUE$
    \ELSIF {$E_1$ \AND $E_2$ \AND $\neg V_{out}$}
        \STATE $E_{out} \leftarrow \TRUE$
    \ENDIF
\ENDFOR
\end{algorithmic}
\end{algorithm}

\begin{figure}[h]
  \centering
  \includegraphics[width=0.7\linewidth]{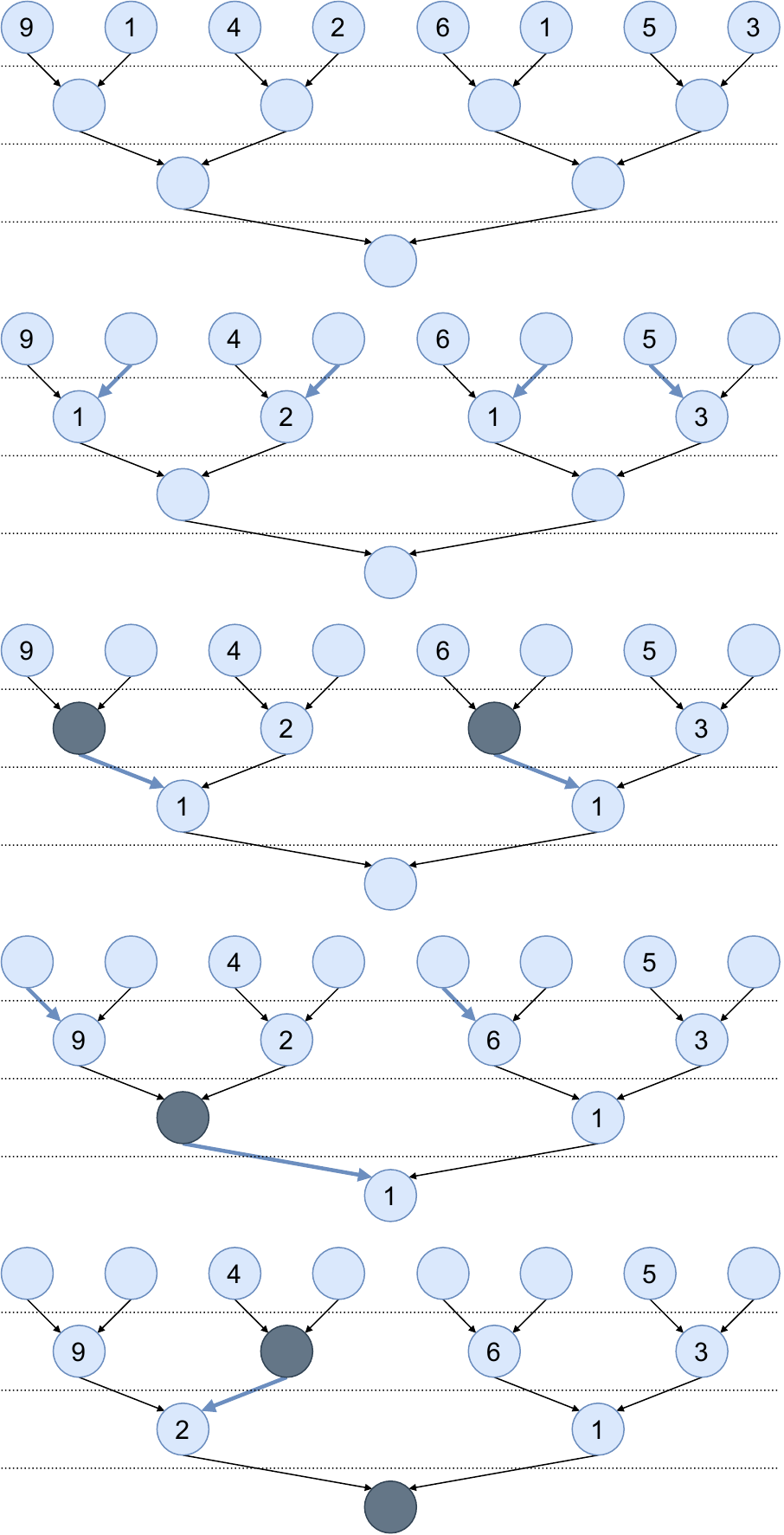}
  \caption{\emph{Bubbling} effect on a simple tree with registers. \emph{Bubbles} are shown in dark, and appear on nodes that are outputting their values, unable to read at the same time.}
  \label{fig:bubbles}
\end{figure}

Note that, because the valid signal in a register can only be modified by the parent when it is false, and by the child when it is true, the critical path no longer travels through more than one comparator, making it constant with respect to $n$. However, this causes an undesired effect where a register can't possibly output two values in a row. We call this effect, illustrated in \Cref{fig:bubbles} \emph{bubbling}, and it renders the output non-streaming after the first value is output in $\mathcal{O}(\operatorname{log}(n))$ cycles.

\emph{Bubbles} appear due to the fact that each register can only be in read or write mode to prevent the critical path from growing with tree depth. In fact, this is the cause of needing the empty signal $E$ at each register, since a parent could be invalid but still have more data lagging behind, in which case we can't make a decision and have to wait. If this effect is looked at at the root of the tree, we find that the output will alternate between valid and empty, for a total sorting time of $\mathcal{O}(2\cdot n)$. For the first $m$ elements, it would take an acceptable $2\cdot m + \log (n)$ cycles.

However, it must be considered that timing is critical \cite{nvidia2025quantumqec} within the context of BP+OSD, and dropping the $2$ from the equation lowers total time from $4m$ to $3m$, an acceleration of $1.33\times$, quite interesting for real-time error correction.



\begin{figure}[h]
  \centering
  \includegraphics[width=0.7\linewidth]{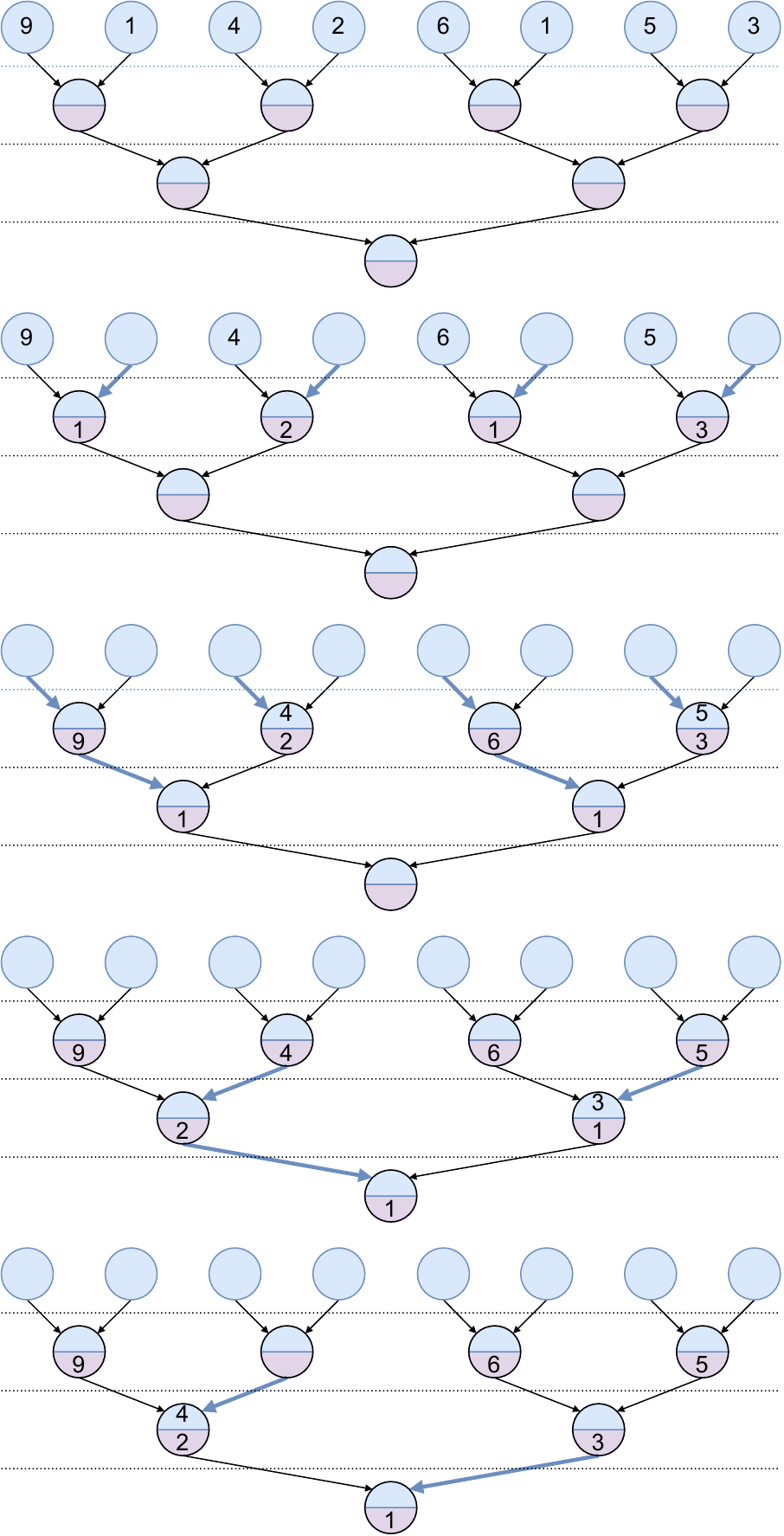}
  \caption{\emph{Bubbling} avoidance by doubling the registers at each node.}
  \label{fig:finalprocess}
\end{figure}

To achieve this goal, a final idea is proposed, named ``hourglass sorting''\footnote{The name is inspired by how the values falling down the tree resemble the grains of sand inside a hourglass}: nodes will have two output registers instead of one, thereby being able to both accept and emit a value each cycle. This idea is commonly known as double or ping-pong buffering \cite{joo1998doubling}. This avoids the creation of \emph{bubbles} while keeping a constant critical path with respect to $n$. How this works is seen in \Cref{fig:finalprocess}, and ensures that after the initial $\mathcal{O}(\operatorname{log}(n))$ initialization cost, the rest of the data streams out sequentially.

The algorithm outline is laid out in \Cref{code:hourglass}. In short, both inputs are compared to decide which one (lowest) moves forward. In case one input is not valid (i.e: no more values will come from that sub-tree) the comparison is ignored. The active input is selected internally to move forward.

The input will be read only if at least a register is empty. Note that, by having two registers, this ensures that we will only prevent writing when both are full, guaranteeing data availability even after a branch is stalled. We always write to register $D_0$ first, then register $D_1$. Thus, the valid signal $V_1\implies V_0$ by also shifting $V_1,D_1$ to $V_0,D_0$ when necessary. $\neg V_1$ can be passed to the input to indicate register availability.

Separately, writes are performed to $D_0$ if it is empty ($\neg V_0$). Note that it is not necessary to check for validity of input $V$ since if $\neg V$, then $\neg V_0$ holds next cycle. When it is full, different decisions are taken depending on if the value is being read by the next node or not. In the first case, the value is replaced by either the input, or register $1$ if it was full. In the second case, register $0$ has nowhere to go so register $1$ is filled with the input.

\begin{algorithm}
\caption{Hourglass cell behavior}
\label{code:hourglass}
\textbf{Input:} Data $D_L$, $D_R$ \\
\textbf{Input:} Valid data flags $V_L$, $V_R$ \\
\textbf{Output:} Ready data flags $V_L$, $V_R$ \\
\textbf{Output:} $D_{out}$; $V_{out}$ \\
\textbf{Input:} $R_{out}$ \\
\textbf{Variables:} $D$, $V$, $R$ \\
\textbf{Registers:} $D_1$, $D_0$, $V_1$, $V_0$ \\

\begin{algorithmic}[1]
\ENSURE Values output in ascending order; sorting is stable.
\FOR{Each cell in the tree, in parallel}
    \STATE $R \leftarrow \neg V_1$
    \IF {$D_L < D_R$} 
        \IF[Select left branch] {$V_L$}
            \STATE $D,V,R_R,R_L\leftarrow D_L,V_L,\FALSE, R$
        \ELSE[Select right branch]
            \STATE $D,V,R_R, R_L  \leftarrow D_R,V_R,R, \FALSE$
        \ENDIF
    \ELSE
        \IF[Select right branch] {$V_R$}
            \STATE $D,V,R_R, R_L  \leftarrow D_R,V_R,R, \FALSE$
        \ELSE[Select left branch]
            \STATE $D,V,R_R, R_L\leftarrow D_L,V_L,\FALSE, R$
        \ENDIF
    \ENDIF
    \STATE $D_{out}, V_{out} \leftarrow D_0, R_0$
    \IF[IN: Fill empty first register]{$\neg V_0$}
        \STATE $D_0, V_0 \leftarrow D, V$ 
    \ELSIF{$\neg V_1$}
        \IF[Simultaneous IN/OUT]{$R_{out}$}
            \STATE $D_0, V_0 \leftarrow D, V$ 
        \ELSE[IN: Fill second register]
            \STATE $D_1, V_1 \leftarrow D, V$ 
        \ENDIF
    \ELSIF[OUT: Shift out values]{$R_{out}$}
            \STATE $D_0, V_0 \leftarrow D_1, V_1$ 
    \ENDIF
\ENDFOR
\end{algorithmic}
\end{algorithm}

Synchronization of both input and output is done via two signals valid ($V$) and ready ($R$) which are asserted by the producer and consumer respectively. When both have asserted their signals, and see the symmetric signal asserted as well, the transaction is assumed completed in that cycle. This means that the produce must empty its register, and the consumer must fill up theirs.

This design achieves our two main goals: First, the input is decoupled from the output, as there is no combinatorial dependency between values or flags going up and down the tree, resulting in no cascading signals. This keeps the critical path under control, and signal fan-out to a minimum. Secondly, \emph{bubbles} can't appear since a node is capable of both inputting and outputting a value in the same cycle.

To prove this, consider the case of a \emph{bubble} appearing at a node. If this was the case, we know both internal registers are empty. If, in the cycle prior, one of the inputs had been available, it would have been shifted into one of the registers since $\neg V_1$ would have held, so both inputs had to be unavailable. This reasoning can be recursively followed up to the leaf nodes, confirming that, in order for a \emph{bubble} to appear, the full tree above must be empty. Therefore it must be the case that the root node is only empty when the full array is sorted. Note that this can only be assured after the first value has been received at the root. This is bound to happen in $\log(n)$ cycles after initialization since, by construction, all of the nodes will have received a value from both parents at that point. Since we can't have \emph{bubbles} afterwards, it follows that the rest of the array is sequentially output in $n$ cycles.



The only thing left is to see that the output is indeed sorted. Consider that, if we assume that both incoming branches for a node are sorting properly, we will always have at our disposal the minimum of both trees. In each case, we select the minimum of the two and pass it on. Neither tree can have \emph{bubbles}, so a value jumping ahead of others, which would disrupt ordering, is also not possible. This not only sorts the tree, but does it in a stable fashion if we give preference to the leftmost sub-tree, which is interesting for many applications. The base case, where a node only has two inputs, will always send the minimum first by construction. (This is, in essence, merge sort with a parallel construction of each level).

\section{Implementation}

\begin{figure}[h]
  \centering
  \includegraphics[width=0.8\linewidth]{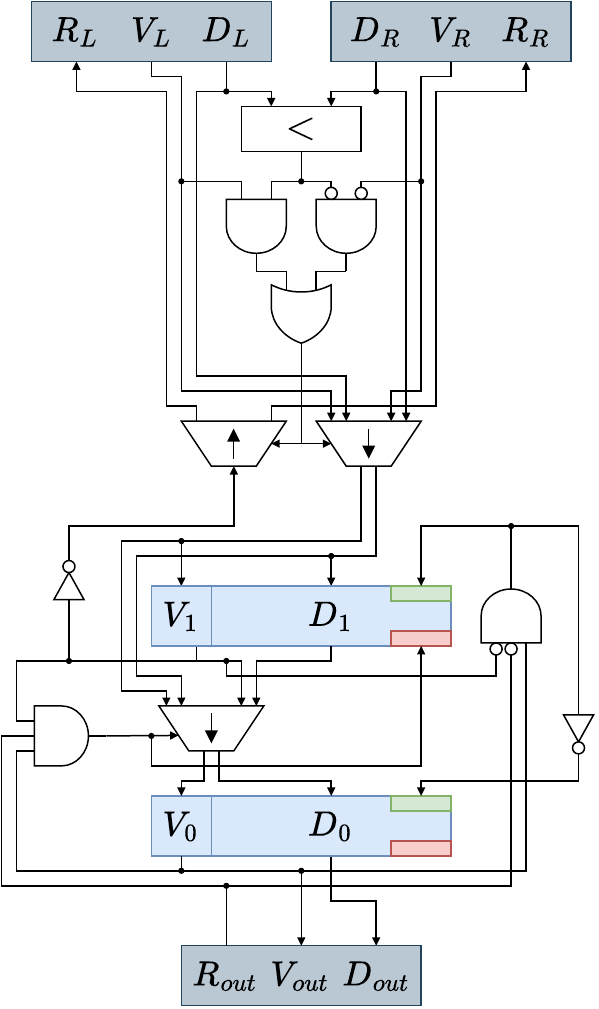}
  \caption{A single sorting cell. In dark blue, the inputs and output, in blue, the registers, with a top enable signal (green) and a bottom clear signal (red). Clock and reset signals are omitted for clarity.}
  \label{fig:sorting_cell}
\end{figure}

A hardware implementation has been done in VHDL and is available on GitHub \cite{hourglasssortingrepo}. The most important modules within the system are the nodes themselves that implement the functionality described in \Cref{sec:idea}. An overview of their implementation is presented in \Cref{fig:sorting_cell}.

Both inputs and outputs are controlled using very lightweight interfaces. They include a data ($D$) signal, along with ready ($R$) and valid ($V$) control bits. A transaction is performed by the endpoints of an interface if and only if valid and ready signals are asserted, respectively, by the sender and receiver.

Following \Cref{code:hourglass}, data is first sorted and selected according to the availability from the input valid signals, and the values themselves. By default, data will flow from the input to register zero (which contains both data $D_0$ and valid $V_0$ registers). In the event that the output is not ready and $V_0$ is asserted, register one will receive the incoming datum instead. By construction, $D_1>D_0$ and $V_1 \implies V_0$. Thus, we always output from register zero, and input when $V_1=0$, therefore $R_{in}=\neg V_1$. To ensure this property holds, register one is ``shifted'' to register zero when both are full, and the output is accepting a new value.

As seen, these operating principles make it so that no paths cross from the output to the input without first going through a register. The critical path is thus extremely fast, spanning just the comparator logic and a few gates and multiplexers.

\begin{figure}[h]
  \centering
  \includegraphics[width=0.8\linewidth]{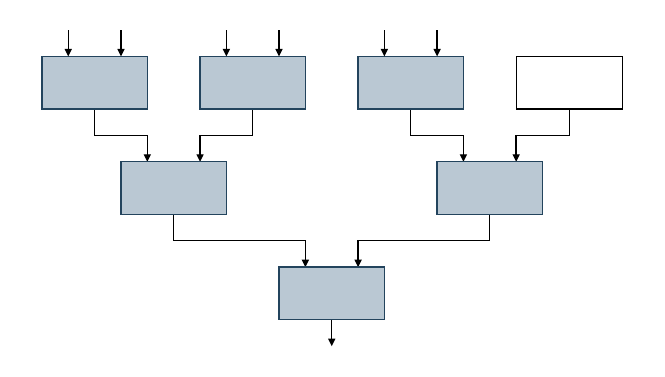}
  \caption{An example of a full tree for $n=6$. Note that the node in layer 2 is not removed even though it only has one parent.}
  \label{fig:sorting_tree}
\end{figure}

To create a sorter for wider arrays, the basic cell is replicated in a tree-like structure (\Cref{fig:sorting_tree}). Special care is taken when a layer of the tree is not a power of two in size, by inserting nodes even when they don't have both parents. This preserves timing integrity, since otherwise values from a sub-tree could jump ahead of others breaking the ordering of the output.

Our implementation can also include index registers, which are used to keep track of the position of the element within the array along with its value (this is not pictured in the diagrams). This is important for the BP+OSD application. Results in this work do not include these index registers, but the overhead in both logic and flip-flop elements would be $n\cdot\operatorname{log}(n)$.

\section{Results}

Various configurations have been synthesized on a xcvu9p-flga2104-2L-e \cite{xcvu9p} device using Vivado \cite{vivado2024}. Values regarding resource use and maximum frequency of operation, with respect to array depth and element width, are presented for various configurations in \Cref{tab:table}. 


\begin{table}[h]
    \centering  
    \begin{tabular}{cccccc}
         Input  & LUT & REG & CARRY8 & Freq & Latency \\\hline\hline 
         1024x8 & 28132 & 27630 & 1023 & 705MHz & 10+1024 \\ 
         512x8 & 14052 & 13806 & 511 & 705MHz & 9+512 \\
         256x8 & 7012 & 6894 & 255 & 705MHz & 8+256 \\
         128x8 & 3492 & 3438 & 127 & 705MHz & 7+128 \\
         64x8 & 1732 & 1710 & 63 & 705MHz & 6+64 \\
         1024x16 & 48592 & 52190 & 1023 & 649MHz & 10+1024 \\ 
         512x16 & 24272 & 26078 & 511 & 649MHz & 9+512 \\
         256x16 & 12112 & 13022 & 255 & 649MHz & 8+256 \\
         128x16 & 6032 & 6494 & 127 & 649MHz & 7+128 \\
         64x16 & 2992 & 3230 & 63 & 649MHz & 6+64 \\ 
         1024x32 & 89002 & 101310 & 2046 & 613MHz & 10+1024 \\ 
         512x32 & 44458 & 50622 & 1022 & 613MHz & 9+512 \\
         256x32 & 22186 & 25278 & 510 & 613MHz  & 8+256 \\
         128x32 & 11050 & 12606 & 254 & 613MHz & 7+128 \\
         64x32 & 5482 & 6270 & 126 & 613MHz & 6+64 \\ 
    \end{tabular}
    \caption{Results for different configurations}
    \label{tab:table}
\end{table}

It is noteworthy to point out that both LUT and REG usage are linearly proportional to $n\cdot w$, where $w$ is the width of each input element. Doubling the number of elements in the array $n$ results in more resources than doubling the width of the elements $w$. Frequency is dependent on element width $w$, which is consistent with the design. Latency is given by $\operatorname{log}(n) + n$, where $\operatorname{log}(n)$ indicates the time to the first output, and $n$ the number of cycles to wait until the array is fully sorted and output.

When compared to other implementations, sorting networks \cite{prasad2011sorting, mueller2012sorting} offer much better latency in the order of $\mathcal{O}(\log^2(n))$, but rapidly run out of resources at costs proportional to $\mathcal{O}(n\cdot \log^2(n))$. Pipelined sorting networks \cite{sklyarov2014high} have the same cost as our proposal, and offer a fixed latency of around $n/2$ for the full array to be sorted, instead of our elastic latency of $\log(n) \to n + \log(n)$ for the first and last elements respectively.


\section{Conclusion}

Sorting algorithms come in all shapes and sizes. In the general case, the limit of $\mathcal{O}(n\cdot \operatorname{log}(n))$ comparisons cannot be lowered, so trade-offs between speed and resource use need to be considered.

Efforts have been made towards fully serial or fully parallel architectures, with a distinct lack of PISO and SIPO approaches in the literature. These are particularly useful when the throughput at the input or output is limited. 

A very simple and lightweight implementation of a PISO sorter is presented. It uses $\mathcal{O}(n)$ resources and is capable of outputting the first sorted element in $\operatorname{log}(n)$ cycles, streaming the rest in exactly $n$. Its frequency does not drop with size, as the critical paths are constrained within single processing elements, which makes it ideal for scalable applications.

The implementation has been successfully integrated within a BP+OSD decoder in the context of quantum LDPC coding, proving its capabilities in real-world applications such as real-time decoding of error-correcting codes.


%

\section*{Acknowledgment}

This work was supported in part by the project PID2023-147059OB-I00 funded by MCIU/ AEI/ 10.13039/501100011033/ FEDER, UE.

\ifCLASSOPTIONcaptionsoff
  \newpage
\fi



%



\bibliographystyle{IEEEtran}
\bibliography{bibtex/bib/IEEEexample}

\begin{thebibliography}{10}
\providecommand{\url}[1]{#1}
\csname url@samestyle\endcsname
\providecommand{\newblock}{\relax}
\providecommand{\bibinfo}[2]{#2}
\providecommand{\BIBentrySTDinterwordspacing}{\spaceskip=0pt\relax}
\providecommand{\BIBentryALTinterwordstretchfactor}{4}
\providecommand{\BIBentryALTinterwordspacing}{\spaceskip=\fontdimen2\font plus
\BIBentryALTinterwordstretchfactor\fontdimen3\font minus
  \fontdimen4\font\relax}
\providecommand{\BIBforeignlanguage}[2]{{%
\expandafter\ifx\csname l@#1\endcsname\relax
\typeout{** WARNING: IEEEtran.bst: No hyphenation pattern has been}%
\typeout{** loaded for the language `#1'. Using the pattern for}%
\typeout{** the default language instead.}%
\else
\language=\csname l@#1\endcsname
\fi
#2}}
\providecommand{\BIBdecl}{\relax}
\BIBdecl

\bibitem{knuth1998art}
D.~E. Knuth, \emph{The Art of Computer Programming: Sorting and Searching,
  volume 3}.\hskip 1em plus 0.5em minus 0.4em\relax Addison-Wesley
  Professional, 1998.

\bibitem{williams1964algorithm}
J.~W.~J. Williams, ``Algorithm 232: heapsort,'' \emph{Communications of the
  ACM}, vol.~7, no.~6, pp. 347--348, 1964.

\bibitem{hollerith1889art}
H.~Hollerith, ``{Art of Compiling Statistics},'' Patent U.S. Patent 395\,781,
  jan 8, 1889, issued January 8, 1889.

\bibitem{hoare1962quicksort}
C.~A. Hoare, ``Quicksort,'' \emph{The computer journal}, vol.~5, no.~1, pp.
  10--16, 1962.

\bibitem{batcher1968sorting}
K.~E. Batcher, ``Sorting networks and their applications,'' in
  \emph{Proceedings of the April 30--May 2, 1968, spring joint computer
  conference}, 1968, pp. 307--314.

\bibitem{ajtai19830}
M.~Ajtai, J.~Koml{\'o}s, and E.~Szemer{\'e}di, ``An 0 (n log n) sorting
  network,'' in \emph{Proceedings of the fifteenth annual ACM symposium on
  Theory of computing}, 1983, pp. 1--9.

\bibitem{cole1988parallel}
R.~Cole, ``Parallel merge sort,'' \emph{SIAM Journal on Computing}, vol.~17,
  no.~4, pp. 770--785, 1988.

\bibitem{mueller2012sorting}
R.~Mueller, J.~Teubner, and G.~Alonso, ``Sorting networks on fpgas,'' \emph{The
  VLDB Journal}, vol.~21, pp. 1--23, 2012.

\bibitem{akl2014parallel}
S.~G. Akl, \emph{Parallel sorting algorithms}.\hskip 1em plus 0.5em minus
  0.4em\relax Academic press, 2014, vol.~12.

\bibitem{sklyarov2014high}
V.~Sklyarov and I.~Skliarova, ``High-performance implementation of regular and
  easily scalable sorting networks on an fpga,'' \emph{Microprocessors and
  Microsystems}, vol.~38, no.~5, pp. 470--484, 2014.

\bibitem{ortiz2010configurable}
J.~Ortiz and D.~Andrews, ``A configurable high-throughput linear sorter
  system,'' in \emph{2010 IEEE International Symposium on Parallel \&
  Distributed Processing, Workshops and Phd Forum (IPDPSW)}.\hskip 1em plus
  0.5em minus 0.4em\relax IEEE, 2010, pp. 1--8.

\bibitem{zuluaga2016streaming}
M.~Zuluaga, P.~Milder, and M.~P{\"u}schel, ``Streaming sorting networks,''
  \emph{ACM Transactions on Design Automation of Electronic Systems (TODAES)},
  vol.~21, no.~4, pp. 1--30, 2016.

\bibitem{jiang2007reliability}
M.~Jiang, C.~Zhao, E.~Xu, and L.~Zhang, ``Reliability-based iterative decoding
  of ldpc codes using likelihood accumulation,'' \emph{IEEE communications
  letters}, vol.~11, no.~8, pp. 677--679, 2007.

\bibitem{terhal2015quantum}
B.~M. Terhal, ``Quantum error correction for quantum memories,'' \emph{Reviews
  of Modern Physics}, vol.~87, no.~2, pp. 307--346, 2015.

\bibitem{peterson1972error}
W.~W. Peterson and E.~J. Weldon, \emph{Error-correcting codes}.\hskip 1em plus
  0.5em minus 0.4em\relax MIT press, 1972.

\bibitem{nvidia2025quantumqec}
\BIBentryALTinterwordspacing
{NVIDIA Quantum Computing Team}, ``Accelerating quantum error correction
  research with nvidia,'' March 2025, accessed: 2025-07-10. [Online].
  Available:
  \url{https://developer.nvidia.com/blog/accelerating-quantum-error-correction-research-with-nvidia-quantum}
\BIBentrySTDinterwordspacing

\bibitem{joo1998doubling}
Y.-M. Joo and N.~McKeown, ``Doubling memory bandwidth for network buffers,'' in
  \emph{Proceedings. IEEE INFOCOM'98, the Conference on Computer
  Communications. Seventeenth Annual Joint Conference of the IEEE Computer and
  Communications Societies. Gateway to the 21st Century (Cat. No. 98},
  vol.~2.\hskip 1em plus 0.5em minus 0.4em\relax IEEE, 1998, pp. 808--815.

\bibitem{hourglasssortingrepo}
\BIBentryALTinterwordspacing
{Daniel Báscones}, ``Hourglass sorting,'' 2025, accessed: 2025-07-10.
  [Online]. Available: \url{https://github.com/Daniel-BG/Hourglass\_sorting}
\BIBentrySTDinterwordspacing

\bibitem{xcvu9p}
\BIBentryALTinterwordspacing
{AMD Xilinx}, \emph{{AMD Virtex UltraScale+ FPGA VCU118 Evaluation Kit -
  Documentation}}, {AMD, Inc.}, 2024, product: XCZU9P-FLGA2104-2L-E. [Online].
  Available: \url{https://www.xilinx.com/products/board-docs/vcu118-docs.html}
\BIBentrySTDinterwordspacing

\bibitem{vivado2024}
\BIBentryALTinterwordspacing
------, \emph{{Vivado Design Suite User Guide}}, {AMD, Inc.}, 2024, version
  2024.1. [Online]. Available:
  \url{https://www.amd.com/en/design-resources/vivado}
\BIBentrySTDinterwordspacing

\bibitem{prasad2011sorting}
D.~Prasad, M.~Y.~M. Yusof, S.~S. Palai, and A.~H. Nawi, ``Sorting networks on
  fpga,'' in \emph{Proceedings of the WSEAS International Conference on
  Telecommunications and Informatics (TELE-INFO)}, 2011, pp. 29--31.

\end{thebibliography}

%
\begin{IEEEbiography}[{\includegraphics[width=1in,height=1.25in,clip,keepaspectratio]{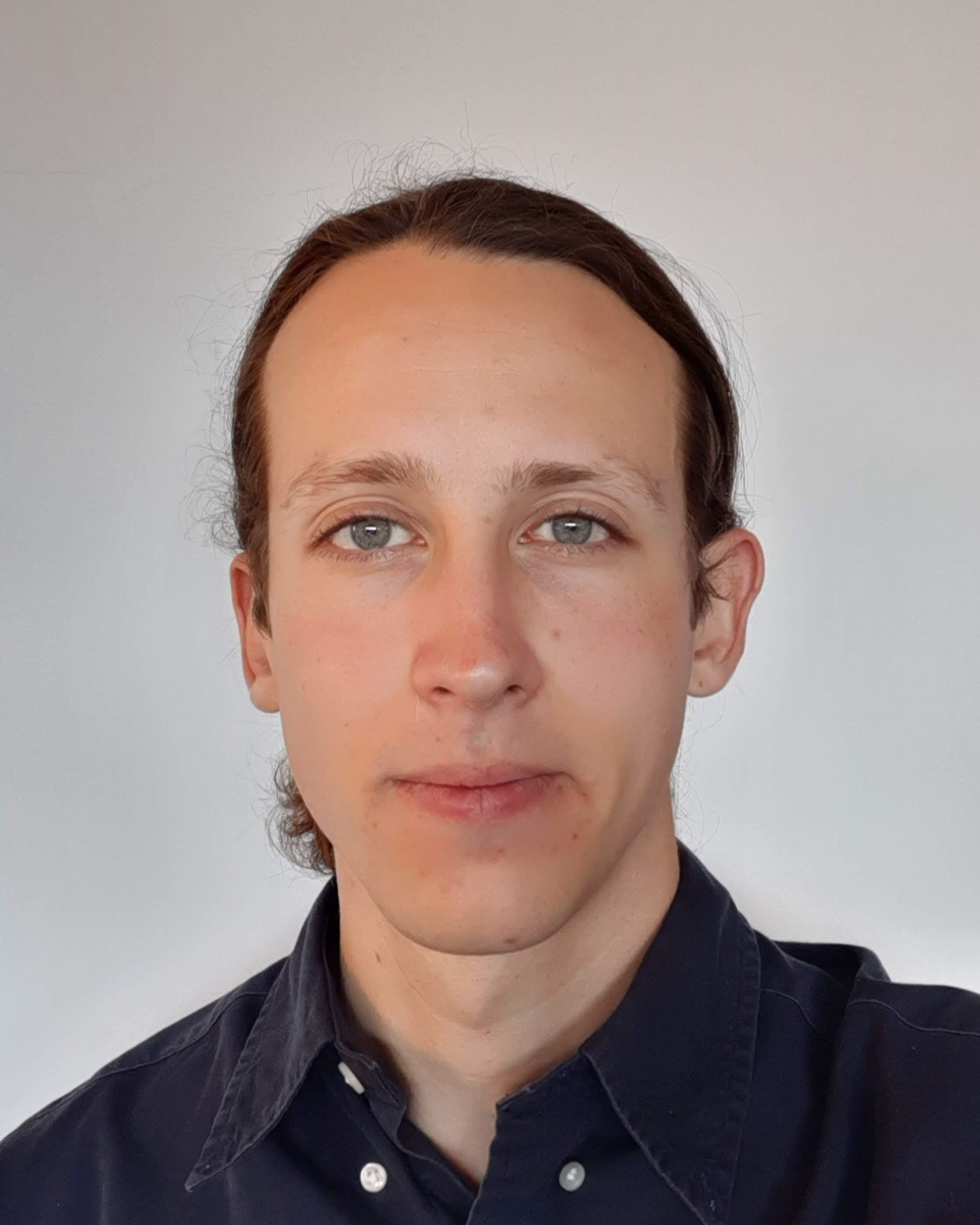}}]{Daniel Báscones}
received the bachelor’s degree in both mathematics and computer science and the M.Sc. degree in computer science from the Complutense University of Madrid, Madrid, Spain, in 2016 and 2018, respectively. He was a Research Associate during the time of his M.Sc. with the Department of Computer Architecture and Automatics. His main interests include hyperspectral image compression on field-programmable gate arrays, dealing with fast lossless algorithms that aid with data transmission and more complex lossy algorithms for long-term storage, a field in which he obtained his Ph.D thesis in 2020.
\end{IEEEbiography}

\begin{IEEEbiography}[{\includegraphics[width=1in,height=1.25in,clip,keepaspectratio]{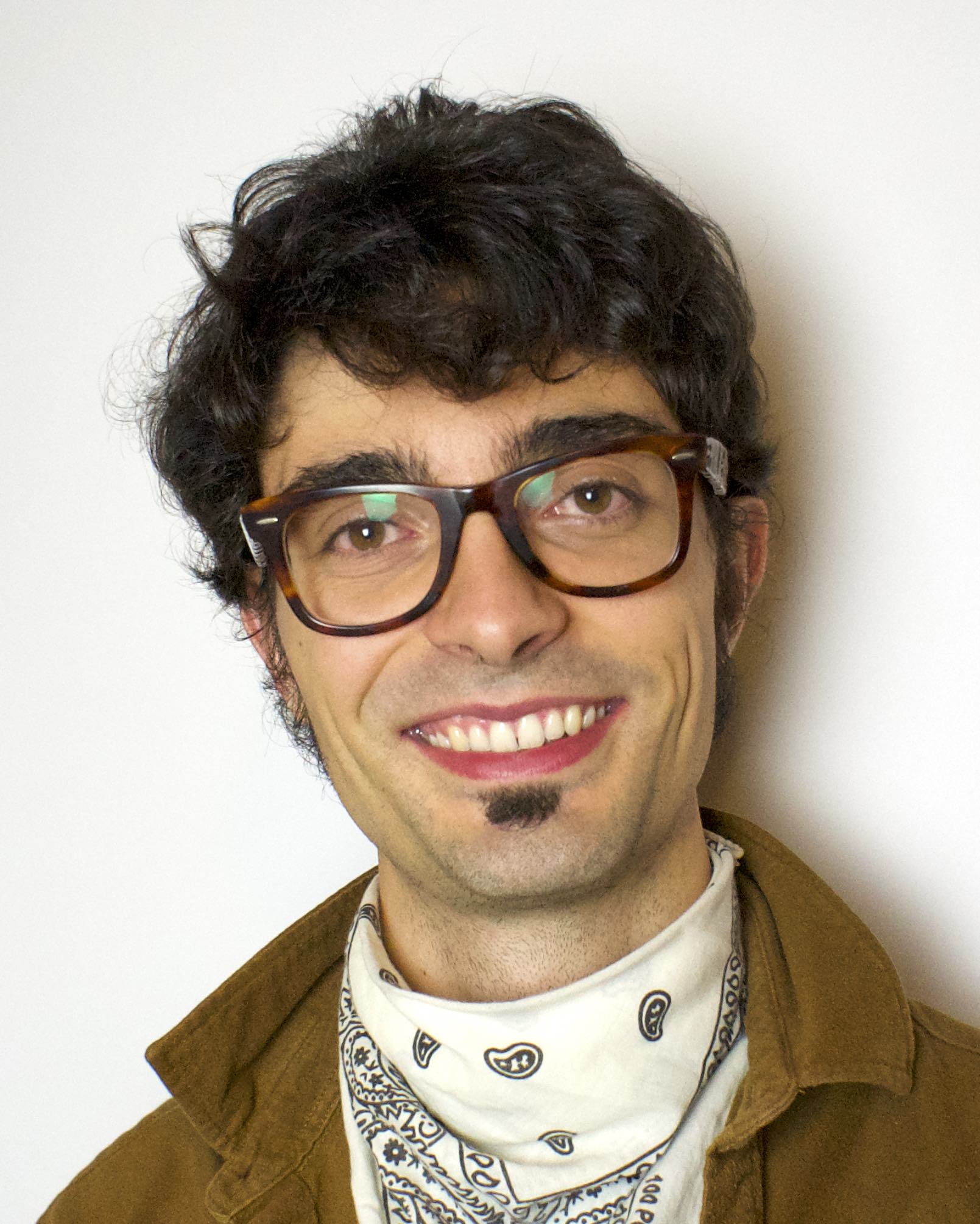}}]{Borja Morcillo}
  received the bachelor and M.Sc. degrees in computer engineering from the Complutense University of Madrid (Madrid, Spain), in 2020 and 2022, respectively. In 2020 he was awarded winner of the Xilinx Open Hardware Design Competition. He is currently a Teaching Assistant (Department of Computer Architecture and Automation, Complutense University of Madrid), while pursuing the Ph.D. degree in hardware design. His main research interests include computer architecture and reconfigurable hardware.
\end{IEEEbiography}




\end{document}